# NUMERICAL INVESTIGATION OF THE VORTEX ROLL-UP FROM A HELICOPTER BLADE-TIP USING A NOVEL FIXED-WING ADAPTATION METHOD

**Antoine Joulain[*], Damien Desvigne[†], David Alfano[‡]**
AIRBUS HELICOPTERS SAS, 13725 Marignane, France

**Thomas Leweke[§]**
CNRS, Aix-Marseille Université, Centrale Marseille, 13384 Marseille, France

### Abstract

This contribution relates to the simulation of the flow around the tip of a helicopter rotor blade in hovering flight conditions. We here propose a new methodology of framework adaptation, using a comprehensive rotor code and high-fidelity numerical simulations. We construct an equivalent fixed-wing configuration from a rotating blade, in which centrifugal and Coriolis forces are neglected. The effect of this approximation on the solution is analysed. The method is validated by a detailed comparison with wind tunnel data from the literature, concerning aerodynamic properties and tip vortex roll-up. This validation also includes variations of the pitch angle and rotational speed, up to transonic tip velocities. Compared to previously published methods of framework adaptation, the new hybrid method is found to reproduce more accurately the flow around a rotating blade tip.

## NOMENCLATURE

| | | | |
|---|---|---|---|
| $(r,\theta,z)$ | = | Rotating-blade frame | |
| $(x,y,z)$ | = | Fixed-wing frame | |
| $(X,Y,Z)$ | = | Inertial frame | |
| $(\delta,n)$ | = | Sectional frame | |
| $(\xi,\eta,\zeta)$ | = | Non-Inertial frame | |
| $c$ | = | Blade chord | m |
| $C_n$ | = | Lift coefficient | - |
| $C_p$ | = | Pressure coefficient | - |
| $C_t$ | = | Thrust coefficient | - |
| $M$ | = | Mach number | - |
| $R$ | = | Blade radius | m |
| $Ro$ | = | Rossby number | - |
| $V_i$ | = | Induced velocity | m.s$^{-1}$ |
| $V_{tip}$ | = | Tip speed | m.s$^{-1}$ |
| $\rho$ | = | Fluid density | kg.m$^{-3}$ |
| $\psi$ | = | Wake age | ° |
| $\psi_S$ | = | Segregation wake age | ° |
| $\Omega$ | = | Rotational speed | rpm |
| CT | = | Caradonna and Tung[1] | |
| KT | = | Kocurek and Tangler[2] | |
| K | = | Komerath *et al.*[3] | |
| S | = | Srinivasan and McCroskey[4] | |

## 1. INTRODUCTION

The region around the blade tip of a helicopter main rotor is associated with complex aerodynamic phenomena[5]. Due to the pressure differential between the upper and lower blade surfaces, a vortical structure is generated in the vicinity of each tip. In hover, the vortices are convected downwards, influencing by induction the loads on all blades of the rotor.

Today's industrial maturity of numerical simulation methods opens the way for a better understanding of the local flow physics in the vicinity of rotor blade tips. The most accurate simulations of isolated rotors use Adaptative Mesh Refinement (AMR)[6], and high-order schemes to reduce numerical diffusion in the wake, as well as high-order turbulence closures or correction for vortical flows (e.g. [7,8]) to avoid unrealistic turbulent diffusion in vortex cores.

However, these strategies are in general too costly to be applied in the frame of tip aerodynamic improvement by numerical design optimization. Low-order schemes and simplified turbulence models are widely used due to the low computational cost. But numerical and turbulent diffusion effects excessively spread out the vortices and reduce their influence on the blade, which leads to inaccurate tip airloads and wake geometry characteristics[9]. Several approaches were proposed in the literature to correct for wake inaccuracy due to diffusion:

- The first and most commonly used strategy is to perform hovering rotor calculations with an approximate source-sink condition (based on Froude's theory) at the outer boundaries[10], in order to compensate the wake-influence deficiency. This method can prevent the emergence of unrealistic flow recirculation when the sensitivity to the far field boundary conditions is significant. However, this method implies an a priori knowledge of the thrust coefficient of the rotor, and shows some sensitivity to the computational domain size. Strawn and Djomehri[11] simulated the experimental configuration of Lorber[12] using a source-sink boundary condition. Comparison revealed a fair agreement of the integrated rotor performance, although the local aerodynamics in the vicinity of the tip was not captured correctly. This is typical of a low-fidelity interaction between the blade and the trailing vortices;

[*]PhD Candidate, antoine.joulain@gmail.com
[†]Engineer, External Aerodynamics, damien.desvigne@airbus.com
[‡]Engineer, External Aerodynamics, david.alfano@airbus.com
[§]Senior Researcher, IRPHE UMR 7342, thomas.leweke@irphe.univ-mrs.fr

- Potsdam et al.[13] performed a weak coupling (transfer of information once every revolution) between a Reynolds-Averaged Navier-Stokes (RANS) code and a comprehensive rotorcraft tool. The coupling strategy consisted in trimming the rotor thrust to the experimental value with the comprehensive tool, and to look for the convergence of the numerical aerodynamic coefficients (i.e. collective pitch angle). The advantage is that the blade deformation can be adjusted at each coupling iteration according to the aerodynamic forces acting on the blade. However, this method cannot correct extra-numerical diffusion, because the local interactional deficiency between the blade and the spread vortices is compensated by a global variation of the collective pitch. Moreover, although the coupling process of Potsdam et al.[13] converged without difficulty, the comparison with measurements showed poor agreement, with an unexpected redistribution of loading from inboard to outboard;

- Various hybrid methodologies involve solving accurate, more or less simplified, Navier-Stokes equations in the vicinity of the blades, and calculating a non-dissipative wake convection in the far field[5]. The first hybridizations were performed with an inviscid potential code in the near field, coupled with a vortex lattice method in the far field[14]. Egolf and Sparks[15] specified the inflow from the vortex lattice wake as a velocity field in the outer boundary of the potential simulation. Remarkable agreement was obtained with the experimental databases of Gray et al.[16] and Caradonna and Tung[1], except in viscous-dominated regions (blade tips and transonic flows). Later, the potential code was replaced by Euler and RANS solvers[17]. The shock strength and position were in better agreement, although some discrepancies were still observed. The mesh refinement, limited by computational resource constraints, was probably the cause;

- Moulton et al.[18], and more recently Bhagwat et al.[19], performed hybrid simulation by using a Thin-Layer Navier-Stokes (TLNS) solver in the vicinity of the blade, and the potential Vortex Embedding (VE) method of Steinhoff and Ramachandran[20] in the far field. Comparison with the measurements of Lorber[12] showed good agreement in the inboard part of the blade, but revealed differences in the vicinity of the tips. The main cause was the lack of validity of the TLNS equations for separated flows;

- The recent Vorticity Transport Model (VTM)[21] uses the vorticity as a conservative variable. The coupling of VTM with a RANS method, performed by Whitehouse and Tadghighi[22], showed promising results in comparison with the experimental data of Caradonna and Tung[1], and emphasizes the need for a comprehensive hover analysis;

- The last strategy is to alter the Navier-Stokes equations in order to confine vorticity and convect vortices without diffusion. The Vorticity Confinement (VC) method of Steinhoff and Underhill[23] involves the addition of an extra term to the momentum equations. The preliminary hover calculations performed by Tsukahara et al.[24], revealed mitigated agreement with the measurements of Caradonna and Tung[1].

In a context related to the numerical optimization of blade tip design, it is important to minimize the cost and complexity of the simulations. One significant simplification consists in considering an equivalent single fixed-wing configuration in a non-rotating Cartesian frame. In this approach, the incoming flow is straight, centrifugal and Coriolis forces are neglected, and the complex geometry at the rotor hub, which is not relevant for the blade tip aerodynamics, is not considered. However, the precise link between rotating and equivalent non-rotating cases, with respect to aerodynamic properties, remains largely an open issue.

A literature survey reveals several attempts to define an adaptation methodology in hover. Srinivasan and McCroskey[4] were the first to look for a numerical strategy to compare the aerodynamics of a rotating blade with an equivalent fixed wing. The base of the method was to keep a constant radial distribution of circulation and to retain the tip Mach number. The sectional circulation $\gamma$ is calculated from the sectional lift $C_n$, the local fluid velocity $V$ and the local blade chord $c$ by:

(1) $\quad \gamma = C_n V c / 2$

Three different ways were proposed to perform a simulation in a fixed-wing configuration:

- The first and most encouraging method is to consider an appropriate Mach function along the span, in order to reproduce the velocity gradient due to the rotation;

- In the second method, a uniform Mach number is kept along the span and the sectional lift coefficient is reduced by an adapted twist distribution;

- The last and most complicated way is to keep a uniform inflow and to alter the local blade chord in order to recover the sectional circulation.

In all methods, the wake influence is taken into account by a constant shift of the pitch angle, whose value is determined by a comprehensive rotor trim[25]. The centrifugal and Coriolis forces are neglected in the fixed wing simulation. In comparison with the measurements of Caradonna and Tung[1], the first adaptation method showed the best agreement in the inboard part of the blade. However, discrepancies were observed near the tip. The simplicity of the employed wake model is probably responsible for the loss of accuracy in the tip area. According to Srinivasan and McCroskey[4], the centrifugal and Coriolis forces appear to have little influence on the tip aerodynamics.

Komerath et al.[3] performed Laser Doppler Velocimetry (LDV) measurements under both rotating and non-rotating conditions, in order to evaluate centrifugal effects on flow separation. The rotating blade was not isolated from its own wake, and the non-uniform flow in the rotating case was not reproduced in the fixed-wing case, which makes this comparison less relevant.

More recently, Vion et al.[26] performed experimental and numerical investigations of a Counter-Rotating Open Rotor (CROR) configuration. The second way of Srinivasan and McCroskey (twist adaptation) was chosen to construct an equivalent fixed-wing configuration, because a velocity gradient cannot be easily reproduced in a wind tunnel. The difficulty of getting an adapted twist law was emphasized and resolved by an iterative process. Numerical comparisons of the rotating and fixed configurations showed good agreement of the tip vortex characteristics, despite a different flow topology in the vicinity of the tip.

We here propose a new methodology of framework adaptation, dedicated to hovering flight. An uncoupled hybrid simulation is set up with a comprehensive rotor code and a high-fidelity Computational Fluid Dynamics (CFD) solver, in order to construct an equivalent fixed-wing configuration from a rotating blade. As for the Egolf and Sparks method[15], the influence of the hover wake is taken into account by a velocity field applied at the boundaries of the CFD domain. The adaptation process is based on Srinivasan and McCroskey's first method[4] and takes into account the induced velocities of the rotating-wing and the fixed-wing wakes. The well-known databases of Caradonna and Tung[1] and Gray et al.[16] are simulated in order to validate the numerical method according to global performance and local tip aerodynamics. The adaptation methodology is presented in detail in Section 2. Results are consolidated in Section 3 by variations of the pitch angle and rotational speed, including a transonic flow case. In Section 4, the new methodology is compared to previously published ones. Finally, the local tip aerodynamics of the rotating wing and the fixed wing are discussed in detail in Section 5.

## 2. NUMERICAL METHODS

### 2.1. Comprehensive Rotor Code

AIRBUS HELICOPTERS' comprehensive rotor code $HOST$[27] is used to trim isolated rotors in hover. The airfoils' aerodynamics is described by two-dimensional (2D) polars obtained from wind-tunnel measurements. The distribution of circulation along the span feeds into a vortex-lattice wake model, while the induced velocities from the vortices are taken into account by a Biot-Savart integration. Results are expressed in the rotating frame in cylindrical coordinates $(r,\theta,z)$. The $r$-, $\theta$- and $z$-axes extend from the root to the tip of the blade, in the direction of rotation and upwards, respectively, the origin being at the center of rotation (Fig. 1).

A trim law imposes fixed values for the collective pitch angle and the rotational speed, whereas the coning angle and thrust coefficient result from a Newton iterative method. All other angular parameters are set to zero.

The prescribed wake model of Kocurek and Tangler (KT)[2], based on the work of Landgrebe[28], was set up using the tip vortex and inboard vortex sheet trajectories of

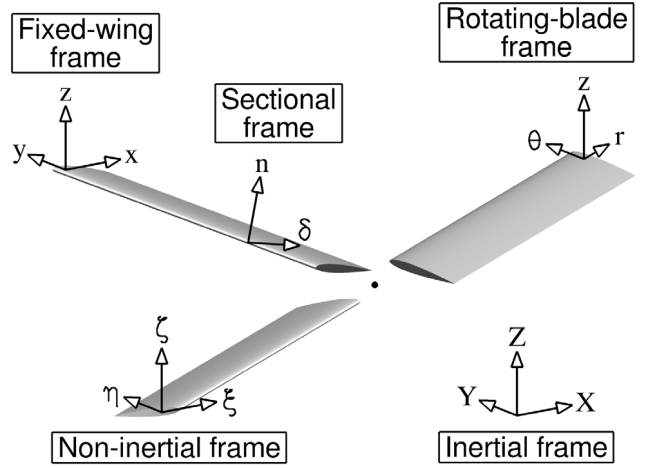

Figure 1. Definition of the inertial $(X,Y,Z)$, the non-inertial $(\xi,\eta,\zeta)$, the rotating-blade $(r,\theta,z)$, the fixed-wing $(x,y,z)$ and the sectional $(\delta,n)$ frames.

26 different rotors. The study included the variation of the chord, the radius, the twist law, the number of blades, the pitch angle and the rotational speed. It should be noted that the influence of sweep, taper, anhedral, the sectional profile and the tip shape has not been investigated. The generalized equations defining the axial $\zeta_t$ and radial $r_t$ tip vortex coordinates as function of the wake age $\psi$ are given by[2]:

(2)
$$\zeta_t = -k_1\psi \text{ for } 0 \leq \psi \leq \psi_b \text{ with } \psi_b = 2\pi/b$$
$$\zeta_t = -k_2\psi + (k_2 - k_1)\psi_b \text{ for } \psi > \psi_b$$
$$r_t = A + (1-A)e^{-\lambda\psi} \text{ with } A = 0.78$$

where $\psi_b$ is the azimuth angle between blades and $b$ the number of blades. The parameters $k_1$ and $k_2$ and $\lambda$ are calculated from

(3)
$$k_1 = B + C(C_t/b^n)^m$$
$$B = -0.000729\Lambda$$
$$C = -2.3 + 0.206\Lambda$$
$$m = 1 - 0.25e^{0.04\Lambda}$$
$$n = 0.5 - 0.0172\Lambda$$

(4)
$$k_2 = -(C_t - C_{t,0})^{1/2}$$
$$C_{t,0} = b^n(-B/C)^{1/m}$$

(5)
$$\lambda = 4\sqrt{C_t}$$

where $\Lambda$ is the blade linear twist. According to Landgrebe[28], it is more convenient to express the axial vortex sheet displacement $\zeta_s$ as a function of the normalized radial location $r$:

(6)
$$\zeta_s = (1-r)\zeta_{s,0} + r\zeta_{s,1}$$
$$\zeta_{s,0} = \begin{cases} 0 \text{ for } 0 \leq \psi \leq \psi_b \text{ with } \psi_b = 2\pi/b \\ -k_3(\psi - \psi_b) \text{ for } \psi > \psi_b \end{cases}$$
$$\zeta_{s,1} = \begin{cases} -k_4\psi \text{ for } 0 \leq \psi \leq \psi_b \text{ with } \psi_b = 2\pi/b \\ -k_5\psi + (k_5 - k_4)\psi_b \text{ for } \psi > \psi_b \end{cases}$$

(7)
$$k_3 = [\Lambda(0.45\Lambda + 18)/128]\sqrt{C_t/2}$$
$$k_4 = -2.2\sqrt{C_t/2}$$
$$k_5 = -2.7\sqrt{C_t/2}$$

Note that the azimuth of the slope variation of $\zeta_{s,0}$ proposed by Landgrebe was $\pi/2$ instead of $\psi_b$. However, the sensitivity to this parameter is expected to be small.

This prescribed model has been widely used since the 1980's for its short computation time (a few minutes) and for its reliability. It is recommended for hover calculations, as long as the rotor characteristics are fully compliant with the framework hypotheses[15].

The vortex lattice method is built upon 2D aerodynamics, under the assumption that the flow field is incompressible, inviscid and irrotational. Thus transonic flow, a three-dimensional (3D) separated boundary layer or tip vortex roll-up cannot be resolved. Several empirical parameters are added to enhance the aerodynamic behavior of the simulation. In particular, the lift coefficient, and thus the circulation, are artificially canceled at the blade tip. Moreover, in order to simulate the tip vortex roll up, the trailing vortex filaments generated between the tip and the radial position of maximum circulation are steadily merged with the tip vortex within several degrees of wake age (Fig. 5). The roll up extent does not significantly influence the numerical results and is taken to be 30°.

## 2.2. Computational Fluid Dynamics Solver

The ONERA *elsA* CFD code[29] is used in this study. The solver is based on a cell-centered, finite-volume approach. Multi-block structured grids are employed to compute the fixed-wing configurations. The application of the Reynolds decomposition to the Navier-Stokes equations leads to the RANS equations, which are written in conservative form:

(8)
$$\frac{\partial \rho}{\partial t} + div(\rho \vec{V}) = 0$$
$$\frac{\partial \rho \vec{V}}{\partial t} + \overrightarrow{div}(\rho \vec{V} \otimes \vec{V}) = -\vec{\nabla}p + \overrightarrow{div}(\bar{\bar{\tau}}) + \vec{f}$$
$$\frac{\partial \rho E}{\partial t} + div[(\rho E + p)\vec{V}] = div(\bar{\bar{\tau}}.\vec{V} - \vec{q}) + \vec{f}.\vec{V}$$

where $\rho$, $\vec{V}$ and $E$ are the mean density, velocity vector and total energy per unit of mass, respectively. The scalar field $p$ is the pressure, $\vec{q}$ the heat flux vector and $\bar{\bar{\tau}}$ the combination of stress and Reynolds tensors. Finally, $\vec{f}$ represents possible source terms. In an inertial reference frame, $(X,Y,Z)$ in Fig. 1, without gravity effects, the source term vanishes. Cast in a non-inertial rotating frame, $(\xi,\eta,\zeta)$ in Fig. 1, the RANS equations bring out two source terms: the centrifugal $(\vec{f_1})$ and the Coriolis $(\vec{f_2})$ forces. The forces $\vec{f_1}$ and $\vec{f_2}$, expressed in Cartesian coordinates, depend on the rotational speed $\Omega$[30]:

(9) $\vec{f_1} = \rho \begin{pmatrix} \Omega^2 \xi \\ \Omega^2 \eta \\ 0 \end{pmatrix}$ ; $\vec{f_2} = \rho \begin{pmatrix} 2\Omega\dot{\eta} \\ -2\Omega\dot{\xi} \\ 0 \end{pmatrix}$

Our objective here is to simulate the flow around a rotating blade in a non-rotating inertial frame, using an appropriately adapted configuration. In such a frame, centrifugal and Coriolis forces are neglected, which implies a free stream without curvature. This new frame is labeled $(x,y,z)$ in Fig. 1. The $x$-, $y$- and $z$-axes extend from upstream to downstream, from the root to the tip of the wing and upwards, respectively, the origin being at the fictitious center of rotation. It can be seen as a snapshot of the rotating frame $(\xi,\eta,\zeta)$ at the time when the $\eta$-axis is aligned with the $r$-axis.

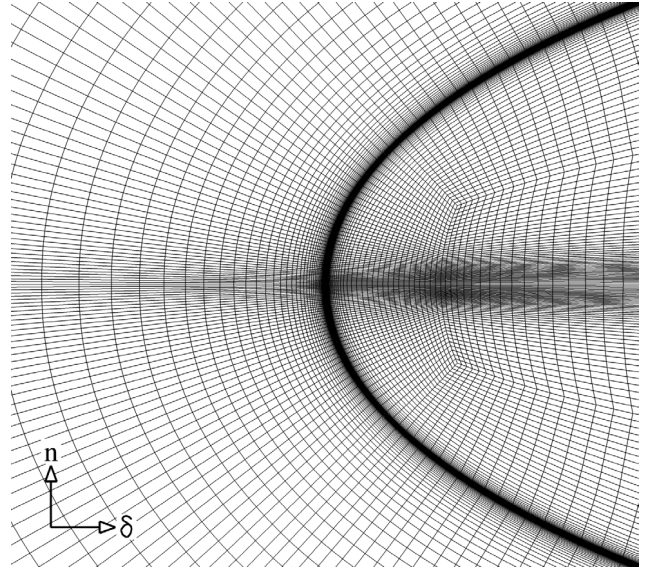

**Figure 2.** Overview of the computational grid in the vicinity of the tip leading edge.

The computational setup is based on a validated numerical method, presented in detail in Ref. [31].

A fully turbulent flow is considered, without boundary-layer transition. The governing equation system is closed with the two-equation $k$-$\omega$ Baseline turbulence model of Menter[32], which is widely used in engineering applications of external turbulent flow prediction. Although this eddy-viscosity model is based on the Boussinesq assumption and is therefore less accurate inside vortex cores[33], it is preferred over second-order closures and corrected models (e.g. [7,8]) for its numerical stability and its short computation time.

The convective fluxes of the mean equations are discretized using the second-order central scheme of Jameson, Schmidt and Turkel[34]. Because this scheme is unconditionally unstable, a scalar artificial dissipation is introduced, as a blend of second- and fourth-order differences. The values of the corresponding dissipation coefficients are taken as 0.5 and 0.032, respectively. The convective fluxes of the turbulent equations are discretized using a second-order Roe scheme. All diffusive flux gradients are calculated with a five-point central scheme. A first-order backward-Euler scheme updates the steady-state solution.

The resulting flow solver is implicit and unconditionally stable, and high values of the Courant-Friedrichs-Lewy (CFL) number can be reached. The CFL number is linearly ramped up from 1 to 10 over 100 iterations, in order to avoid divergence during the transient phase.

The present simulations concern a rectangular fixed wing placed in a straight incoming flow, intended to represent one blade of a rotor.

The structured multiblock grids are designed with the ANSYS ICEM-CFD software. A 2D C-H topology surrounds the profile and the downstream zone. The 3D mesh is constructed by stacking 2D meshes along the span. Beyond the tip, the resulting gap is filled with a half-butterfly mesh (O-grid mesh). The domain dimensions extend 200 chords in all directions from the tip. In the wall normal direction, the first grid spacing is chosen in order to fix the dimensionless wall distance ($y^+$, see e.g. Tennekes and Lumley[35]) between 0.7 and 1. Expansion ratios of the grid spacings are not greater than 1.15. An overview of the mesh in the vicinity of the tip leading edge is shown in Fig. 2. A typical mesh is constituted of roughly 130 blocks and 20 million points.

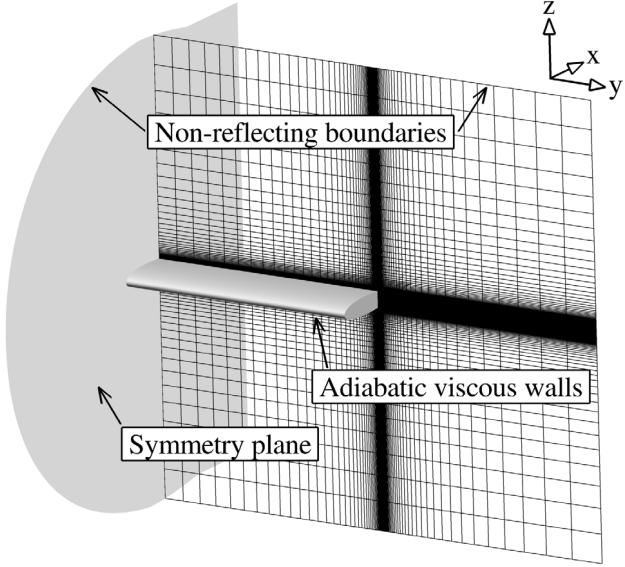

Figure 3. Boundary conditions of the CFD fixed-wing computations.

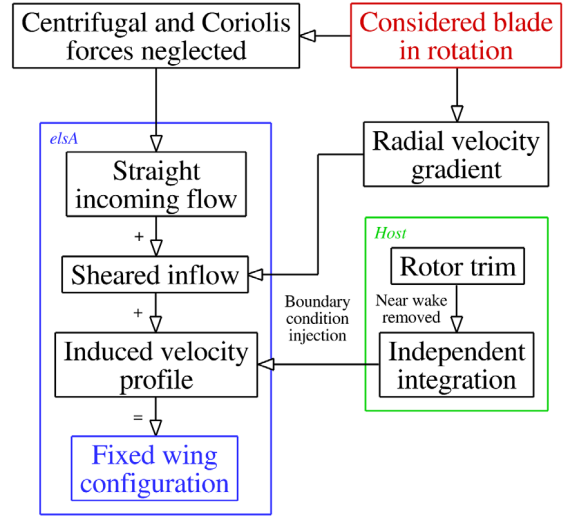

Figure 4. Overview of the fixed-wing adaptation process.

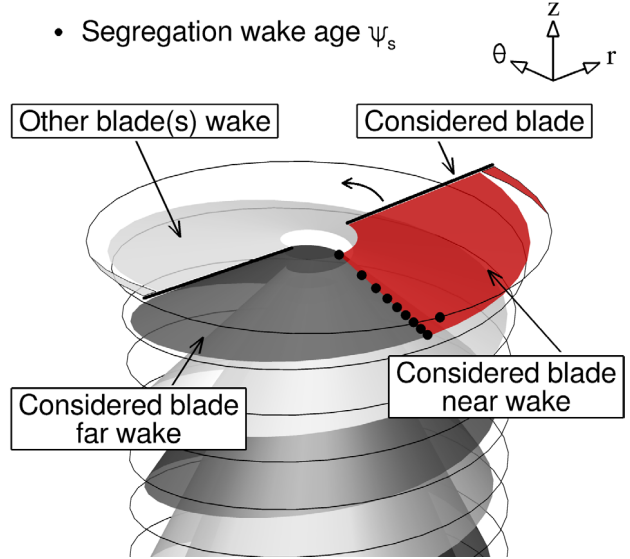

Figure 5. Definition of the segregation wake age $\psi_s$, the near and far wakes of the considered blade and the wake of the other blade(s).

The wing surface is modeled as an adiabatic viscous wall (zero heat-flux). The boundary supporting the root of the wing is a symmetry plane, whereas a non-reflecting condition is applied to all other far-field boundaries (Fig. 3).

Approximately 20,000 iterations (20 hours on 48 processors) are required to obtain a converged solution with a 4-orders-of-magnitude decrease of the $L^2$ norm-based residual of the total energy.

## 2.3. Hybridization procedure

On the one hand, the RANS simulations are able to accurately compute 3D separated boundary layers and tip vortex roll-up, including the influence of viscosity and compressibility. However, the numerical dissipation spreads out the vortices too quickly and degrades their influence on the blade, which leads to inaccurate rotor airloads and wake geometry characteristics.

On the other hand, wake properties are quickly estimated using the comprehensive rotor code. Tip vortices are analytically propagated over a long distance without dissipation, resulting in a good qualitative prediction of the wake influence on the blades. However, the simplified aerodynamic model implemented in the tool is too restrictive to study the tip vortex roll-up in detail.

The adaptation process presented here takes advantage of both tools and proceeds in three steps, as illustrated in Fig. 4. First, the rotating blade is represented by a fixed wing in a rectangular domain. Centrifugal and Coriolis forces are neglected, thus the fixed-wing is exposed to a straight incomming flow. The main features linked to the rotation are taken into account by a specific choice of the initial and boundary conditions.

Following Srinivasan and McCroskey[4], the radial gradient of relative velocity of the rotating blade is represented by a sheared inflow along the span of the wing ($V_x = \Omega r$), as shown by the symbols in Fig. 8a. In order to avoid numerical instabilities, a minimum velocity has to be imposed in the computational area in the vicinity of the centre of rotation, its value was chosen as $V_x = \Omega c/2$ for most simulations. In the outer part of the domain, the inflow velocity was limited to a constant maximum value $\Omega(R + 2c)$. The stability of the computation and the solution are not sensitive to this parameter.

Finally, the influence of the rotor wake is taken into account in the fixed-wing computation by the injection of an appropriate induced velocity profile at the boundaries of the domain. It is obtained from the three-dimensional induced velocity at the location of the considered blade, computed using the comprehensive rotor code *HOST*.

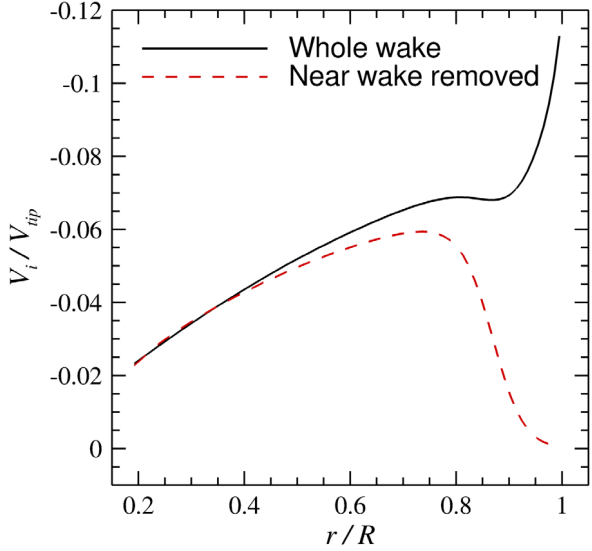

**Figure 6.** Induced velocity profiles integrated from the whole wake and with the considered blade near wake removed.

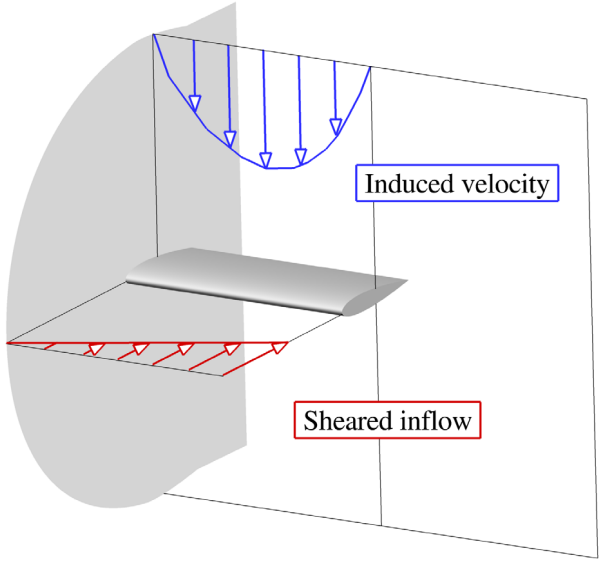

**Figure 7.** Illustration of the velocity profiles injected at the boudaries of the fixed-wing domain.

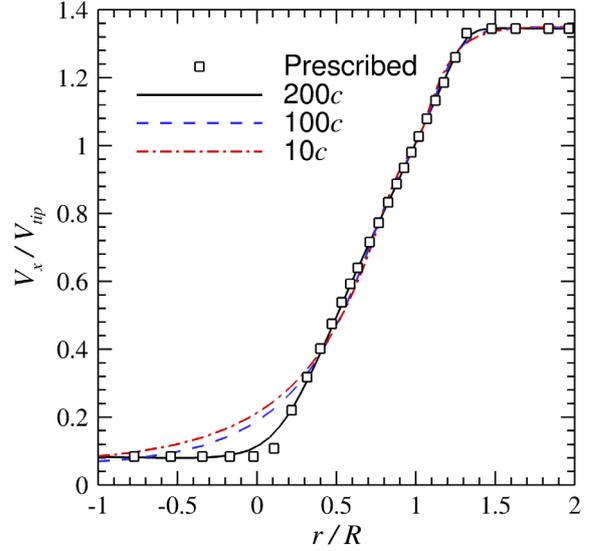

a)

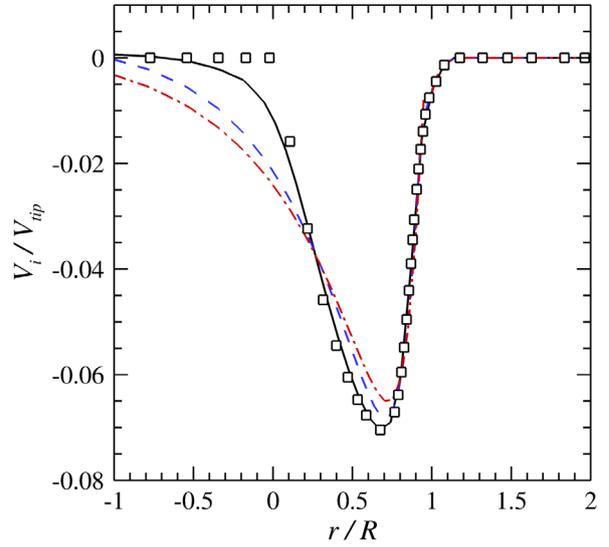

b)

**Figure 8.** Prescribed distribution and hybrid-CFD velocity profiles at $200c$, $100c$ and $10c$ above the blade: a) inflow ; b) induced velocity.

*HOST* is used to trim the rotor with the Kocurek and Tangler[2] (KT) wake model. The wake is calculated over 15 revolutions, in order to minimize the dependence of the wake length on the trim. However, the fixed-wing generates its own wake, which also alters the velocity distribution along the span of the wing. In order to avoid accounting twice for the near-wake influence, an additional step is performed. At a certain wake age $\psi_s$, the wake sheet generated by the considered blade is segregated into a near wake and a far wake (Fig. 5). An independent Biot-Savart integration is performed in each part to compute the distribution of the vertical induced velocity $V_i$ (along $z$-axis) on the considered blade. Only contributions from the far wake of the considered blade and the wake(s) of the other blade(s) are accounted for. The missing contribution from the near wake is provided by the CFD simulation, which is far more accurate in this region. The segregation wake age $\psi_s$ can be chosen between 30° and 90° without significant modification of the solution on the blade. In fact, in this interval, the vortex is far from the considered blade and its influence decays proportionally to the square of the distance.

Figure 6 compares the induced velocity profiles on the considered blade, integrated from the whole wake and with the considered blade near wake removed. As expected, the influence of the near wake is maximum at the tip and is quickly reduced to zero far from $r/R = 1$.

The CFD tool is then used to construct the fixed-wing equivalent configuration. The sheared inflow and the induced velocity distribution, extracted on the blade from *HOST*, are injected as far field boundary conditions in the CFD simulation (Fig. 7). These profiles (as functions of span $y$) are invariant from upstream to downstream and from the top to the bottom of the computational domain, which is justified since the principle aim is to correctly represent the flow in the blade tip region. No induced velocities are injected in the axial ($x$) and radial ($y$) directions, because these components are small compared to the vertical one, and in order to avoid problems related to continuity. The velocity profiles are also used to initialise the whole interior of the computational domain at the beginning of the simulation.

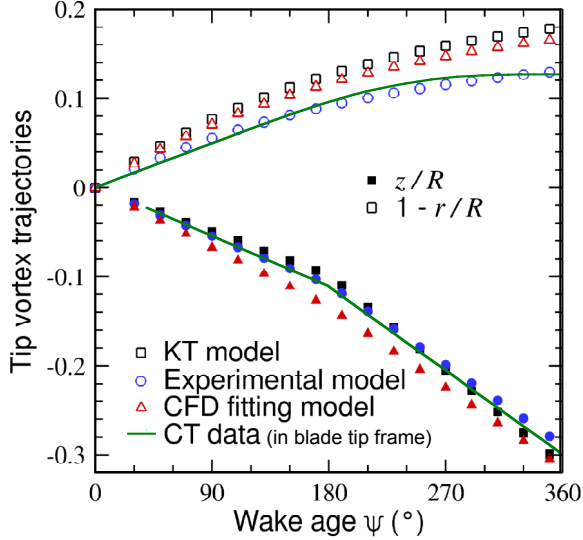

**Figure 9.** Axial and radial tip vortex trajectory. HOST calculations with different wake models and experimental data from Caradonna and Tung[1].

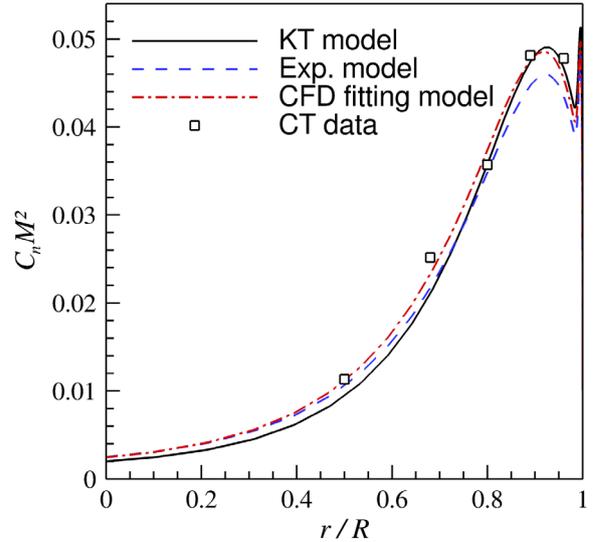

**Figure 10.** Normalized sectional lift coefficient along the span. Hybrid CFD simulations with different wake models and experimental data from Caradonna and Tung[1].

In Fig. 8, the normalized velocity profiles resulting from the calculation are plotted for several locations above the blade. In comparison with the prescribed distributions, the numerical profiles are in good agreement in the vicinity of the tip ($r/R = 1$). Some discrepancies appear around the center of rotation ($r/R = 0$). In this area, the large cell dimension increases the numerical dissipation, thus the high velocity gradients are smoothed.

The size of the computational downstream domain does not influence the solution in the range [$10c$-$200c$], i.e. blade vortex ageing from a quarter to five revolutions. In fact, the wing trailing vortex follows a straight trajectory, thus only the near wake contributes to the induced velocities on the tip.

## 3. TEST CASE OF CARADONNA AND TUNG[1]

The well-known experimental database of Caradonna and Tung[1] is chosen to perform a validation of the methodology. The configuration consisted of two blades with a constant, untwisted and untapered NACA 0012 airfoil. The chord ($c$) and the radius ($R$) were respectively 0.191 m (7.5 in.) and 1.143 m (45 in.). The internal radius of the blade (first aerodynamic profile) was not given, so it is arbitrarly fixed to one chord. Moreover, the tip cap geometry is assumed to be flat. This rotor is essentially rigid, thus no flexible model is needed. The blades were instrumented with pressure taps between $0.5R$ and $0.96R$. This measurement resolution does not allow a comprehensive study of the tip vortex itself, but is useful to validate the hybrid numerical calculation. A good agreement of the sectional lift profiles along the blade span is an indication of an accurate interaction between the rotor and the wake. Given the scarcity of pressure measurements in regions with high pressure gradients (leading edge, shock waves) in the Caradonna and Tung[1] database, the sectional drag profiles are not calculated from this data and not compared to the numerical results.

It has to be mentioned that, in various tables of Ref. [1], some values of the integrated sectional lift are inverted. For example, the test case characterized by a pitch angle of 8° and a rotational speed of 2500 rpm is summarized in Table 25 and plotted in Fig. 6 of Ref. [1]. The last three radial stations are clearly inverted. A trapezoidal rule integration, performed from the published pressure measurements, reveals that the figure is correct. All data from Caradonna and Tung[1] used in this study have been corrected.

### 3.1. Reference Test Case

The reference test case is characterized by a collective pitch angle of 8°, a rotational speed of 1250 rpm and a thrust coefficient of 0.00459. The tip Reynolds and Mach numbers are 1.94 million and 0.436, respectively.

The Kocurek and Tangler[2] (KT) model is first used to perform the hybrid simulation. The trajectory of the tip vortex in the $z$- and $r$-directions, as calculated by *HOST*, is plotted as a function of the wake age in Fig. 9 (square symbols). At $\psi = 180°$, the tip vortex is estimated to convect vertically a distance $0.098R$ below the rotor plane and a distance $0.126R$ radially inward.

The lift coefficient in the sectional frame ($C_n$), resulting from the application of the hybrid procedure, is shown in Fig. 10. A normalization is applied by multiplying $C_n$ by the square of the local Mach number $M$ in order to account for the influence of the sheared inflow. The numerical result is in good agreement with the measurements. The amplitude and location of the maximum lift, in the vicinity of the tip ($r/R \approx 0.9$) is correctly reproduced. In the inboard part of the blade, the computation slightly underpredicts the lift coefficient. The pressure coefficient $C_p$, plotted in Fig. 11 as a function of the nondimensional chord location $\delta/c$ for $r/R = 0.96$, confirms the high accuracy of the hybrid procedure.

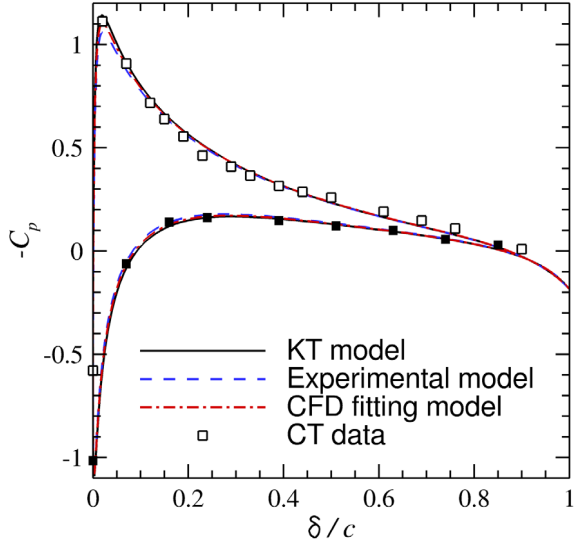

Figure 11. Sectional pressure coefficient at $r/R = 0.96$. Hybrid CFD simulations with different wake models and experimental data from Caradonna and Tung[1].

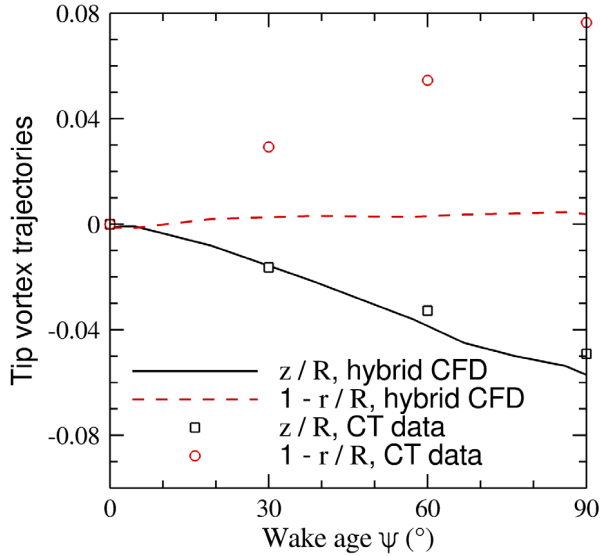

Figure 12. Axial and radial tip vortex trajectory. Hybrid CFD simulation with KT model and experimental data from Caradonna and Tung[1].

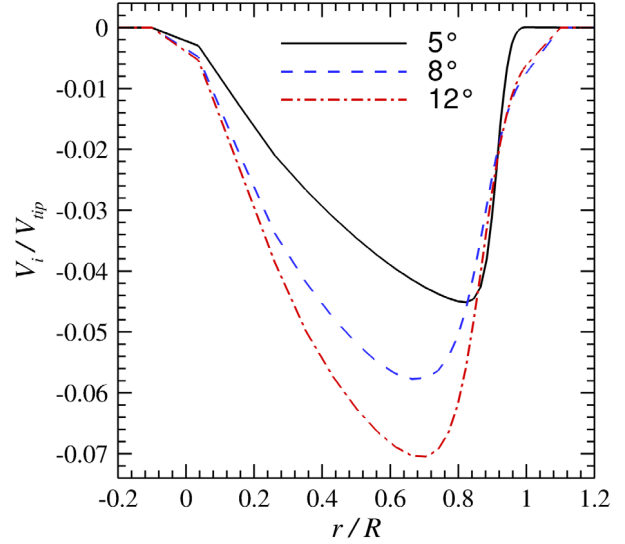

Figure 13. Induced velocity distributions calculated from *HOST* for different collective pitch angles.

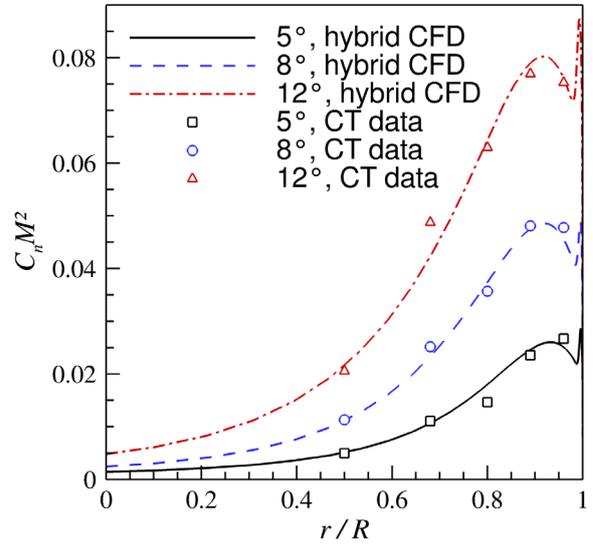

Figure 14. Sectional lift coefficient distributions for different collective pitch angles. Hybrid CFD simulations and experimental data from Caradonna and Tung[1].

The trajectory of the vortex resulting from the hybrid CFD simulation, obtained from the local maxima of the $Q$-function[36], is compared to the experimental data of Caradonna and Tung[1] in Fig. 12. Concerning the axial trajectory, good agreement is obtained, which indicates that the total induced velocity (integrated from *HOST* for the far wake and computed by CFD for the near wake) is correct. Since no induced velocity is injected in the radial direction in the CFD simulation, the location of the tip vortex along the $r$-axis is virtually constant, the radial contraction is not reproduced. However, this is not detrimental for the computation of the global aerodynamics of the blade, that are very little influenced by the "older" parts of the tip vortex ($\psi \geq 30°$).

In order to assess the influence of the *HOST* wake model on the CFD solution, different vortex trajectory laws are compared. An approximate model is constructed from the experimental trajectory by modifying the coefficients of the KT model according to:

(10) $\begin{aligned} A &= 0.796 \\ C &= -3.135 + 0.206\Lambda \end{aligned}$

At $\psi = 180°$, the measured radial and axial vortex locations of wake age are respected. In comparison with the KT model, the experimental model leads to a very similar axial convection, but a decrease of the radial contraction (Fig. 9). At $\psi = 180°$, the latter is reduced from $0.126R$ to $0.092R$ (-30%). As a consequence, the tip vortex from the preceding blade is closer to the considered blade tip and a redistribution of induced velocity occurs around $r/R \approx 0.8$. Inboard of $r/R \approx 0.8$, the induced velocity is lower, while it is higher outboard. The sectional lift coefficient (Fig. 10) and the pressure coefficient (Fig. 11) reveal a substantial deterioration (by more than 6%) of the solution in the vicinity of the tip leading edge. Concerning the inboard part of the blade, the experimental wake model is in better agreement with the data.

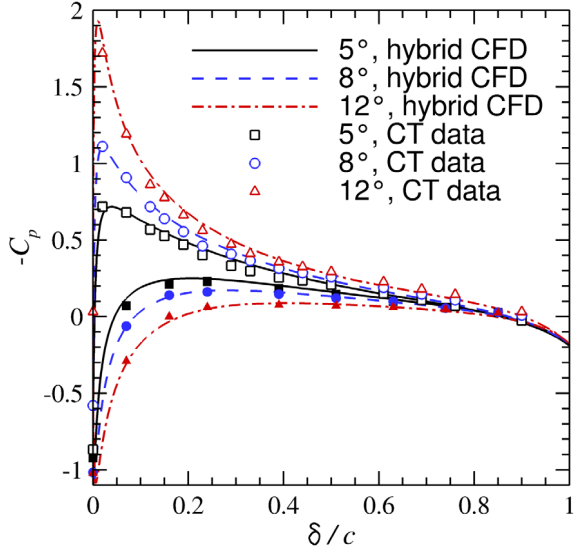

a)

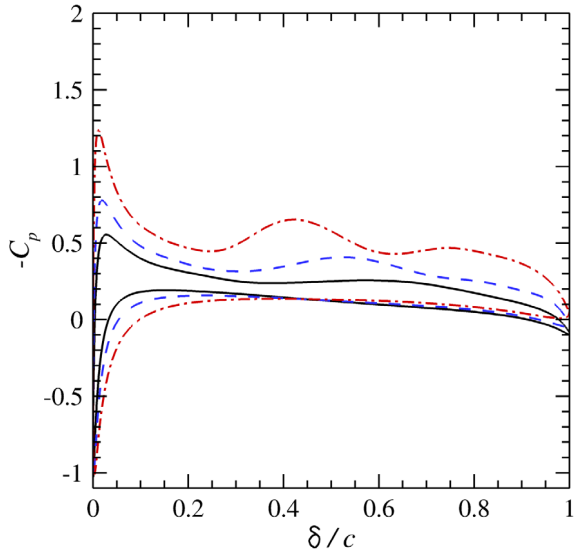

b)

**Figure 15.** Sectional pressure coefficient for different collective pitch angles. Hybrid CFD simulations and experimental data from Caradonna and Tung[1]:
a) $r/R = 0.96$ ; b) $r/R = 0.995$.

Caradonna and Tung[1] indicated that the discrepancy between the KT model and the measurements can be explained by measurement error. Here, a CFD fitting model is constructed from several combinations of radial and axial trajectory perturbations. At $\psi = 180°$, the best result is obtained with an increase of the vertical convection by 36% and a decrease of the radial contraction by 7% (Fig. 9). As a consequence, the lift coefficient in the inboard part is slightly increased and fits the measurements (Fig. 10), without altering the tip aerodynamics (Fig. 11). To conclude, the CFD fitting model is conserved for the reference test case.

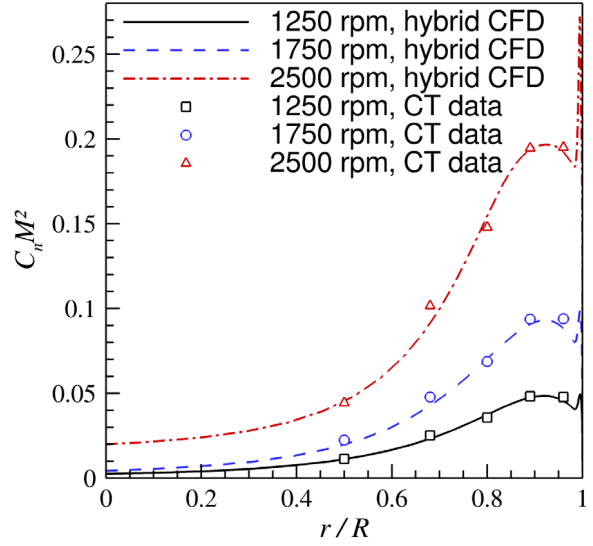

**Figure 16.** Sectional lift coefficient distributions for different rotational speeds. Hybrid CFD simulations and experimental data from Caradonna and Tung[1].

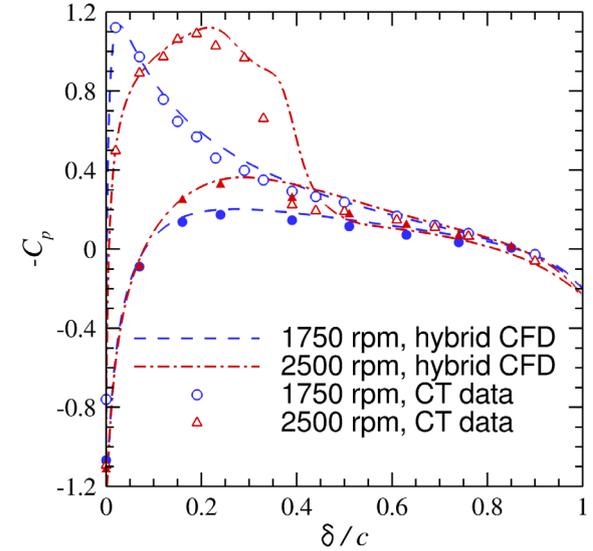

**Figure 17.** Sectional pressure coefficient at $r/R = 0.96$ for different rotational speeds. Hybrid CFD simulations and experimental data from Caradonna and Tung[1].

### 3.2. Variation of Pitch Angle

The Caradonna and Tung[1] database allows a study of the influence of the collective pitch angle on the numerical hybrid solution. Given a rotational speed of 1250 rpm, the pitch angle is reduced to 5° (thrust coefficient of 0.00213) and increased to 12° (thrust coefficient of 0.00796). For these two new cases (and those of the next section), the time-consuming construction of a CFD fitting wake model is not carried out. Instead, they are calculated with the experimental wake model, which gives very similar results.

As expected, the induced velocity increases with the pitch angle (Fig. 13). With respect to the 5° case, the maximum amplitude of the induced velocity at 8° and 12° is increased by 28% and 56%, respectively. The radial location of the maximum slightly moves from approximately 0.8 at 5° to 0.75 at 8° and 12°.

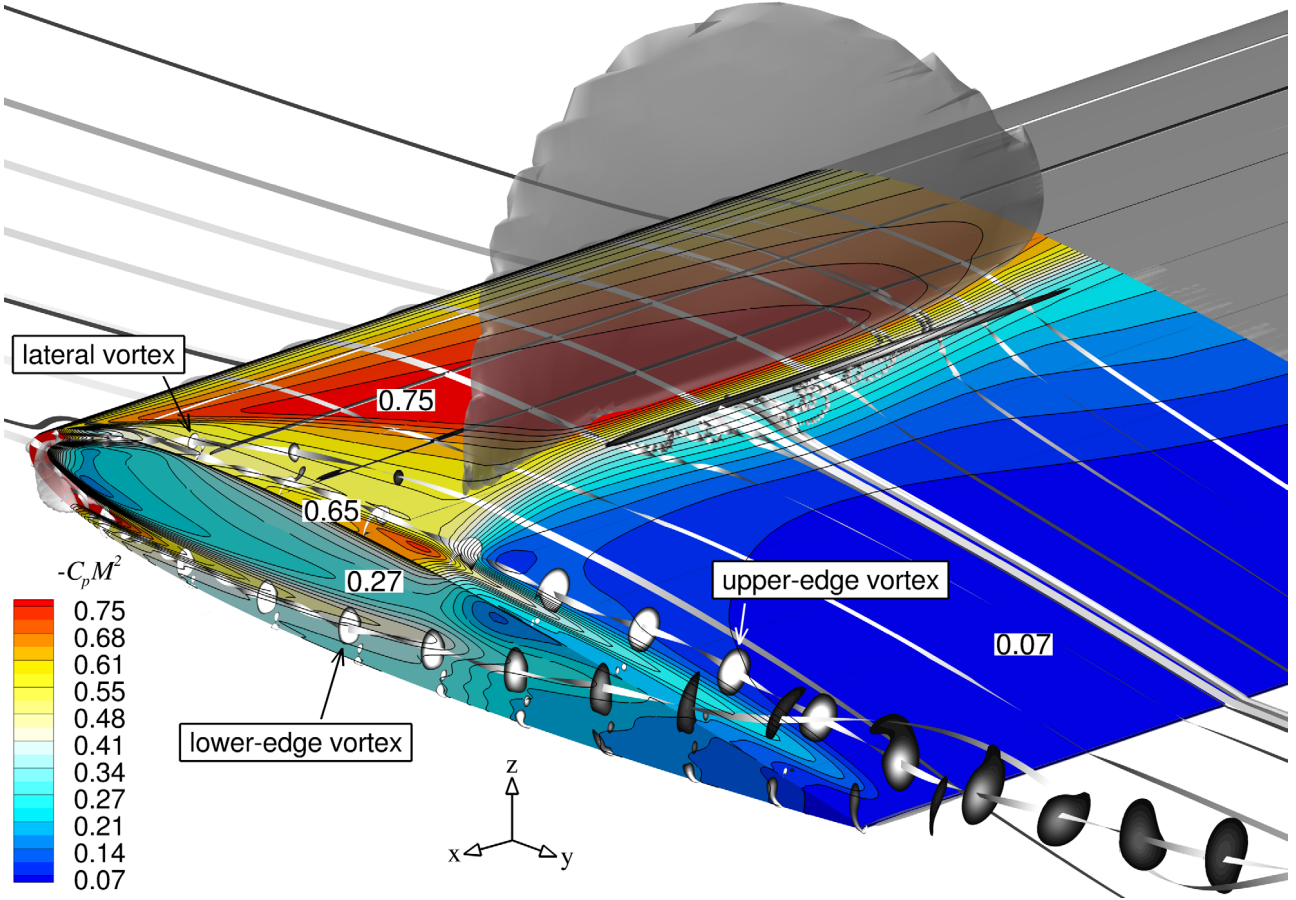

**Figure 18.** Hybrid CFD simulation for 8° pitch angle and 2500 rpm rotational speed. Normalized contours of the pressure coefficient; positive $Q$-function in cross-sectional planes between $\delta/c = 0$ and 1.4; isosurface of high velocity gradient along the $x$-axis; and streamlines.

The solutions of the hybrid procedure are compared to the measurements in Figs. 14 and 15a. Very good agreement is obtained for all pitch angles and all radial locations. At $r/R = 0.96$, the efficiency of the hybrid computation is confirmed by the comparison of the pressure profiles. Up to this radial location, the flow is essentially 2D due to the moderate pitch angles and the rotational speed.

The accuracy of the CFD tool allows a comprehensive study of the vortex roll-up in the vicinity of the tip. Figure 15b reveals that the pressure at $r/R = 0.995$ is strongly altered by the pitch angle. At 5°, one vortex suction peak, of very low amplitude is visible on the upper side at $x/\delta \approx 0.65$. At 8°, this peak increases in amplitude and moves to $x/\delta \approx 0.55$. Moreover, a second suction peak of small amplitude is visible at $x/\delta \approx 0.80$. Finally, the 12° case shows an increase of the amplitude of both suction peaks, while their locations are closer to the leading edge.

### 3.3. Variation of Rotational Speed

The rotational speed was varied at 8° of collective angle. The value of $\Omega$ is increased from 1250 rpm to 1750 rpm and 2500 rpm. The corresponding tip Mach numbers are 0.436, 0.607 and 0.877, respectively. The thrust coefficient is not sensitive to the rotational speed.

The 1250 rpm case is calculated with the CFD fitting wake model (Section 3.1), whereas the experimental wake models are used for the two higher rotational speeds.

Due to the inception of a numerical instability in the region of low velocity, the minimum inflow value for $V_x$ is increased from $\Omega c/2$ to $\Omega c$ for the 2500 rpm test case.

The sensitivity of the normalized induced velocity to the rotational speed is low. With respect to the 1250 rpm case, the maximum amplitude of the induced velocity at 1750 rpm and 2500 rpm is increased by 6% and 14%, respectively. Again, the sectional lift distributions along the radius computed with the hybrid procedure are in very good agreement with the measurements (Fig. 16). With respect to the 1250 rpm test case, the local maximum localised at $r/R \approx 0.9$ is multiplied by a factor 2 and a factor 4 at 1750 rpm and 2500 rpm, respectively. The lift amplitude in the vortex footprint, visible at $r/R \approx 0.995$ is also multiplied by a factor 2 between 1250 rpm and 1750 rpm. However, a factor 5.5 is noted between 1250 rpm and 2500 rpm, which indicates the presence of a non-linearity in the flow field.

In the two first measurement sections of the 2500 rpm case, i.e. $r/R = 0.5$ and $r/R = 0.68$ (as well as for all radial stations of the 1750 rpm case, including $r/R = 0.8$), a closely 2D flow is observed, and the previous conclusions concerning the accuracy of the hybrid procedure apply. However, the three last measurement sections exhibit a strong discontinuity on the upper surface. This shock is visible on the pressure coefficient profile at $r/R = 0.96$ in Fig. 17. Numerically, the amplitude of the shock is well reproduced, but its location is slightly shifted downstream (by $0.05c$) in comparison with the measurements.

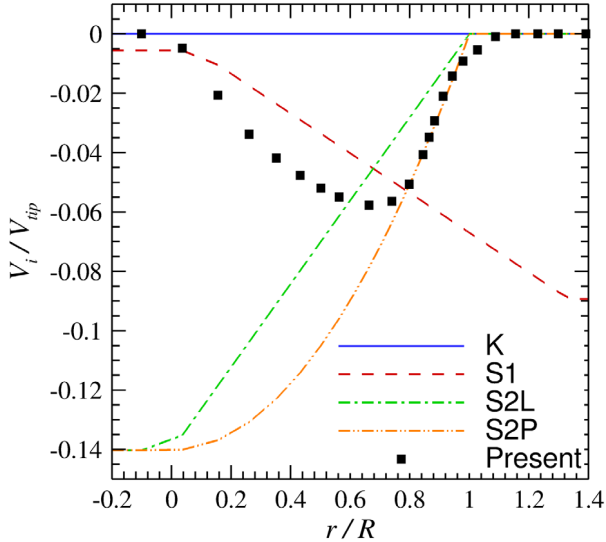

**Figure 19.** Induced velocity distributions with the K, S1, S2L and S2P strategies compared to the present adaptation method.

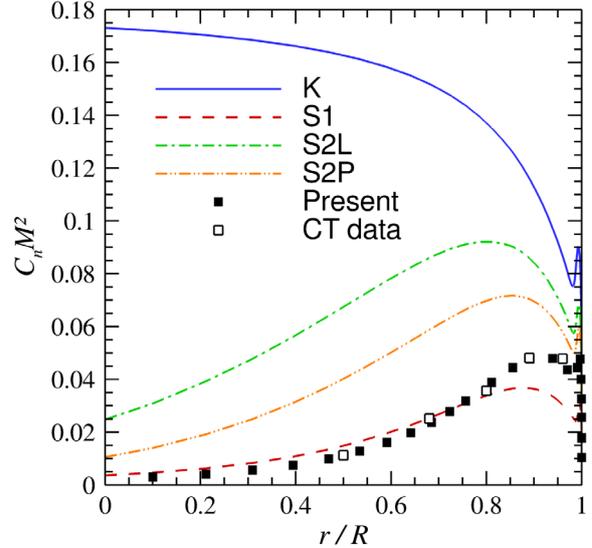

**Figure 20.** Sectional lift coefficient distributions obtained with different strategies compared to experimental data from Caradonna and Tung[1].

This transonic flow case exhibits complex aerodynamic phenomena. The accuracy of the numerical tool allows a comprehensive analysis of the tip vortex roll-up; a detailed investigation is presented in a separate publication [31]. As an example, Fig. 18 shows the numerical solution in the region within one chord of the tip. The normalized pressure coefficient $C_p M^2$ is plotted as color contours. In the vicinity of the tip, the vortical flow has a high influence on the pressure field on the upper surface and on the lateral side.

Positive values of the $Q$-function[36] are shown in cross-sectional planes between $\delta/c = 0$ and 1.4 in steps of $0.1c$. Two strong vortices are generated by the two sharp edges of the truncated geometry. The lower-edge vortex moves up along the side of the tip. It is strained and wrapped around the upper-edge vortex further downstream. The proximity of the vortices to the blade induces a suction force near the vortex paths, which manifests itself as a local increase of $-C_p M^2$. In the vicinity of the tip leading edge, a small lateral vortex is generated along the end cap. This vortex is driven inboard and quickly dissipated under the influence of the strong upper-edge vortex.

The isosurface of high velocity gradient along the $x$-axis highlights the shock location. On the upper surface, the normal shock is interacting with the boundary layers and the 3D tip aerodynamics. A small oblique shock wave is observed in the vicinity of the lateral leading edge.

The streamlines reveal that a boundary layer separation occurs right after the upper-surface shock. In the vicinity of the tip, the vortex roll-up reduces the shock intensity and prevents the separation of the boundary layer.

### 4. COMPARISON TO PREVIOUSLY PUBLISHED METHODOLOGIES

In this section, the hybrid adaptation methodology is compared to the various other strategies identified in the literature survey. The reference test case of Caradonna and Tung[1] is simulated (collective pitch angle of 8°, rotational speed of 1250 rpm).

The fixed-wing approach presented by Komerath et al.[3] does not take into account wake considerations, neither the non-uniform rotational inflow. Thus, the fixed configuration with the K strategy is described by a uniform Mach 0.436 inflow, no induced velocity (Fig. 19), and a constant pitch angle of 8°. The first method of Srinivasan and McCroskey[4] (S1 strategy) is implemented here by choosing an appropriate Mach function along the span to reproduce the velocity gradient due to the rotation. A constant shift of the pitch angle (-3.8°) is applied for the entire fixed blade, which is equivalent to the linear induced velocity profile shown in Fig. 19.

The second way of Srinivasan and McCroskey[4] is to keep a uniform Mach number along the span and to reduce the sectional lift coefficient by an adapted twist distribution. Since the optimal twist profile is unknown, a linear evolution is first considered (S2L). Then, as proposed by Vion et al.[26], a parabolic law is evaluated (S2P).

The induced velocity profiles from the above methods are potted in Fig. 19. The S2P profile is very close to the present one between $0.8R$ and the tip.

The sectional lift coefficient distributions obtained with the five strategies are plotted in Fig. 20. Due to the absence of induced velocity, the K strategy clearly overestimates the sectional angle of attack and lift coefficient. Moreover, the slope of the curve is not reproduced because of the constant Mach number along the span. The S1 strategy reveals a good behaviour from the centre of rotation to $0.7R$. In this region, the induced velocity profile is close to the present one. From $0.7R$ to the tip, the induced velocity profile and the lift coefficient show differences. The sectional lift distributions are overestimated by the S2 strategies, although the S2P method is close to the experimental result in the vicinity of the tip.

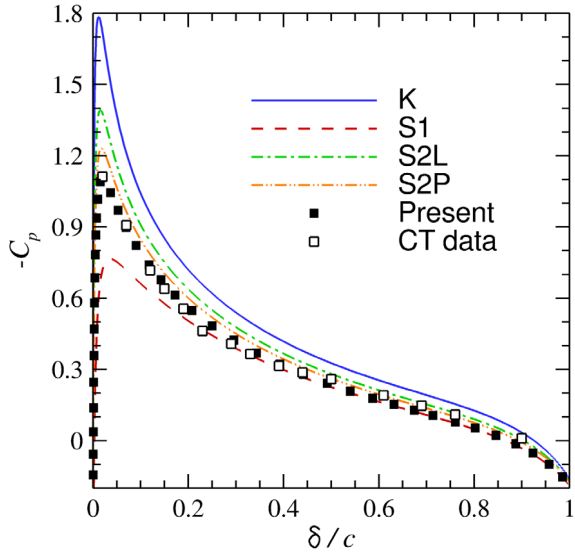

a)

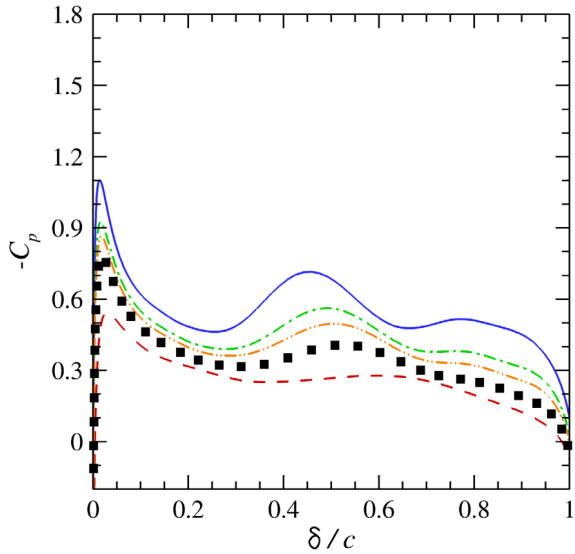

b)

**Figure 21.** Sectional pressure coefficient on the upper surface obtained with various strategies. a) $r/R = 0.96$ ; b) $r/R = 0.995$.

This is confirmed by a comparison of the sectional pressure coefficient at $0.96R$ (Fig. 21a). With respect to the measurement, the leading-edge suction peak is overestimated by 60%, 25% and 11% with K, S1 and S2P, respectively, whereas it is underestimated by 30% with S2L. Finally, the sectional pressure coefficients across the vortex footprint ($r/R = 0.995$) are plotted in Fig. 21b. The amplitude of the vortex suction peaks is highly altered by the adaptation strategy. With respect to the present method, the S2P adaptation deviates by 22% at $\delta/c \approx 0.5$.

A better agreement could probably be obtained with the S2 strategy by performing an iterative optimization of the twist law (in the inboard part of the blade), as emphasized

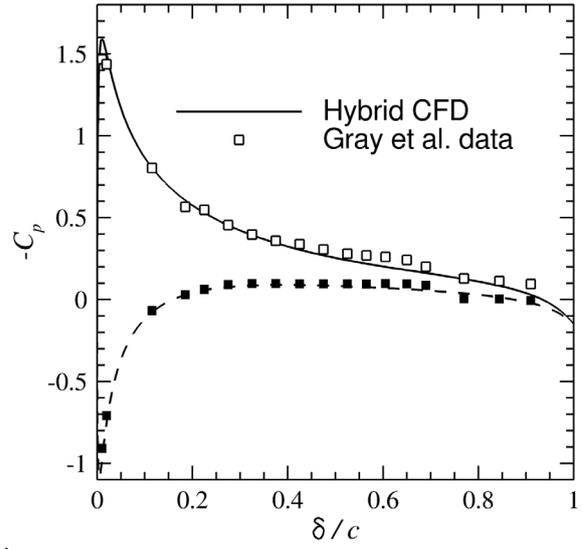

a)

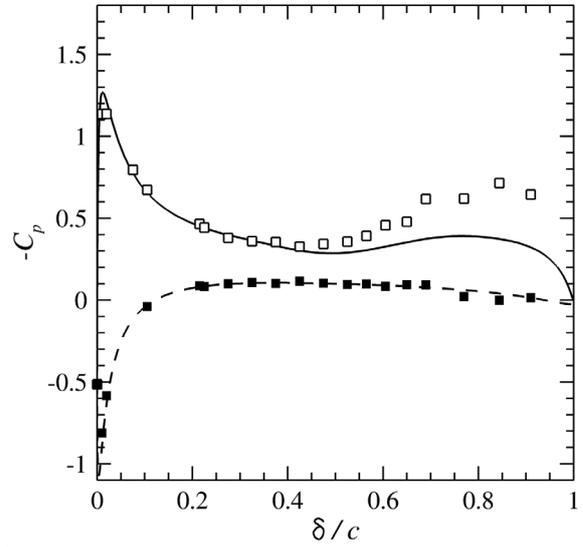

b)

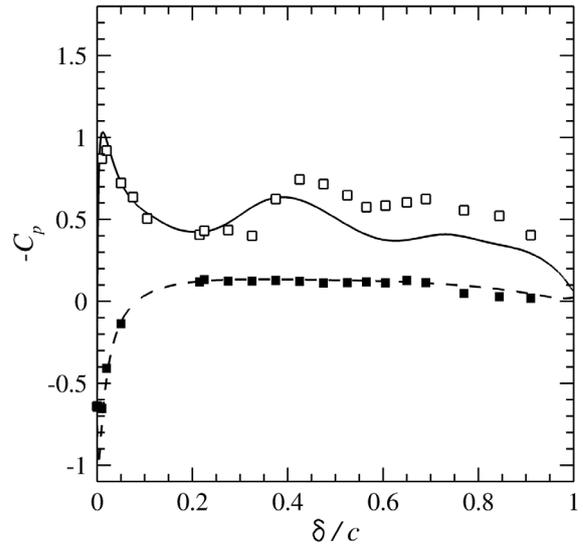

c)

**Figure 22.** Sectional pressure coefficient. Hybrid CFD simulations and experimental data from Gray *et al.*[16]: a) $r/R = 0.966$ ; b) $r/R = 0.987$ ; c) $r/R = 0.995$.

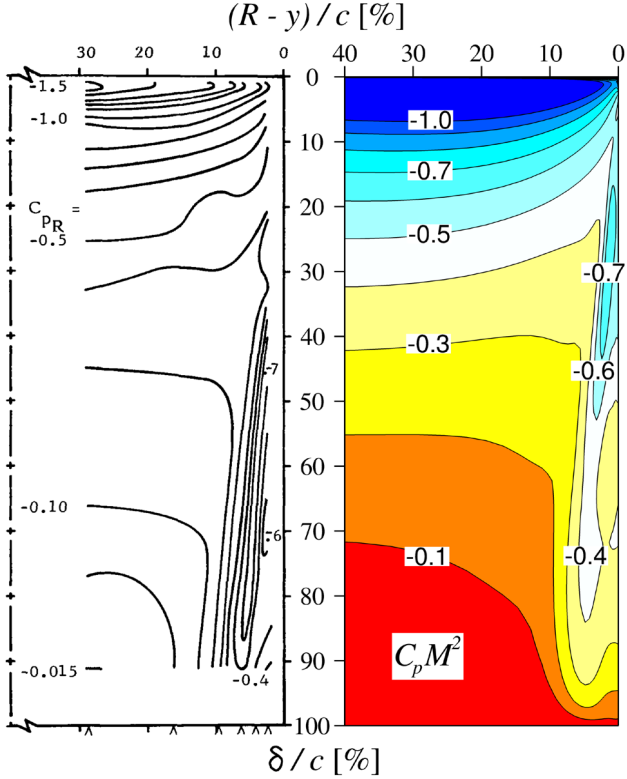

Figure 23. Contours of constant pressure coefficient (normalized by the tip Mach number) on the upper surface. Hybrid CFD simulation (right) and experimental data extracted from Gray et al.[16] (left).

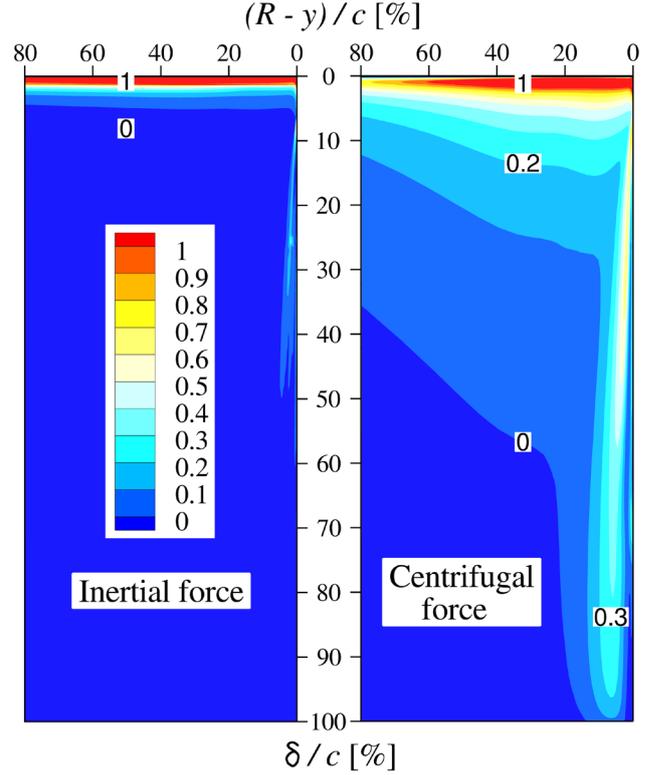

Figure 24. Contours of the normlized inertial and (velocity-affecting) centrifugal forces in the upper-surface boundary layer, calculated a posteriori from the hybrid CFD solution.

by Vion et al.[26]. However, this process is time consuming and was not investigated further.

## 5. TEST CASE OF GRAY ET AL.[16]

The last simulated test case is extracted from the Gray et al.[16] database. A single-bladed rotor is used to investigate in detail the local influence of the tip vortex roll-up on the blade pressure. The blade is made of an untwisted NACA 0012 airfoil, with a constant chord of 0.127 m (5 in.), a radius of 0.61 m (24 in.) and a truncated tip. The rotational speed and the pitch angle are fixed to 1350 rpm and 11.4°, respectively. The resulting chord-based Reynolds and Mach numbers at the tip are 0.74 million and 0.25.

The *HOST* trim is performed using the KT model. Based on the Caradonna and Tung[1] database, the thrust coefficient is estimated as 0.006. This value is confirmed a posteriori with the Hybrid CFD solution.

The sectional pressure coefficient at $r/R = 0.966$ (Fig. 22a) shows good agreement between the hybrid CFD simulation and the measurements. At this radius, the vortex roll-up does not influence the pressure, and the flow is two-dimensional.

Figures 22b and 22c show results for $r/R = 0.987$ and 0.995, in the vortex footprint. The pressure at the leading-edge suction peak and on the whole lower surface is well predicted. However, discrepancies appear on the vertical suction peak. The amplitude of the local maximum is underestimated by 45% and shifted upstream by $0.07c$ at $r/R = 0.987$. At $r/R = 0.995$, the two vortical peaks identified in the measurements are numerically reproduced, but deviations of the amplitude and location are again seen.

In order to understand the origin of these deviations, contours of the normalized pressure coefficient on the upper surface near the tip are plotted in Fig. 23. Two observations can be made: (1) the pressure decrease in the path of the numerical and experimental vortices is of the same amplitude, but the former is shifted upstream by approximately $0.15c$ with respect to the latter, i.e. the numerical vortex generated from the lateral upper edge starts to roll up before the experimental one. (2) Despite this delay, the radial position of the experimental vortex path in the vicinity of the trailing edge is slightly more inboard (by $0.02c$ at $\delta/c = 0.9$), i.e. the radial displacement of the experimental vortex is larger than the numerical one.

Several simplifications were made in the construction of the present adaptation method. In particular, the radial induced velocities of the far wake are not taken into account. According to the *HOST* simulation, this radial velocity is directed inboard and is maximum at $0.8R$. A one-chord vortex age results in an inboard displacement of $0.027c$, which is of the same order of magnitude as the radial displacement deviation.

The centrifugal and Coriolis forces are neglected in the present hybrid strategy. Under the assumption of a high blade aspect ratio, the centrifugal force at the tip is aligned with the $r$-axis. Its expression in the fixed-wing $(x,y,z)$ frame (Fig. 1) derives from Eq. 9 and is $\vec{f_1} = \rho \Omega^2 r \vec{y}$. This force can be written as the sum of two terms:

$$(11) \quad \rho\Omega^2 r\vec{y} = \vec{\nabla}\left(\rho\Omega^2 \frac{r^2}{2}\right) - \Omega^2 \frac{r^2}{2}\vec{\nabla}\rho$$

The first one derives from a potential flow, and can be included in the pressure gradient term of Eq. 8. The second term represents the part of the centrifugal force acting on the velocity field.

The norms of the inertial force and the velocity-affecting centrifugal term ($\|\overrightarrow{div}(\rho\vec{V}\otimes\vec{V})\|$ and $\|\Omega^2 r^2 \vec{\nabla}\rho/2\|$, respectively) are compared in the upper-surface boundary layer near the tip in Fig. 24. These distributions are calculated a posteriori from the hybrid CFD solution. In the vicinity of the tip leading edge and in the path of the upper-edge vortex, the centrifugal force is higher than the inertial force. As a consequence, in the case of a rotating blade, the flow in this area is probably radially ejected from the tip under the influence of the centrifugal force. In particular, the small lateral vortex generated in the vicinity of the tip leading edge (observed in Fig. 18) is likely to be stronger than in the fixed-wing adapted simulation, since it results from the streamwise vorticity of the blade boundary layer related to radial outflow, delaying the generation of the upper-edge vortex. Thus the accelerated roll-up in the present simulations is likely related to the absence of centrifugal effects.

The influence of the Coriolis force can be estimated by calculating the Rossby number. This number represents the ratio between the inertial force (driving the flow field) and the Coriolis force, and can be expressed as:

$$(10) \quad Ro = \frac{\text{Inertial force}}{\text{Coriolis force}} \sim \frac{V}{2\Omega c}$$

In the vicinity of the tip, assuming that the order of magnitude of the velocity is $V = \Omega R$, the Rossby number is proportional to the aspect ratio of the blade $R/c$. For this particular test case, the Rossby number is equal to 2.4, which suggests that the Coriolis force may influence the flow field. However, the norm of the Coriolis force is several orders of amplitude lower than the norm of the centrifugal force in Fig. 24. Moreover, the aspect ratio of classical main rotors is higher than in the present configuration. The influence of the Coriolis force is therefore expected to be very low and can be neglected in the study of the tip vortex roll-up.

## 6. CONCLUSIONS

A new methodology of framework adaptation is presented for hovering flight. A fixed-wing equivalent configuration is constructed from a rotating blade, in order to simplify and speed up tip vortex simulations. An uncoupled hybrid strategy is set up using the comprehensive rotor code *HOST*[27] and the high-fidelity CFD solver *elsA*[29]. The former is used to correctly account for the induction of the far wake by propagating the vortices over a long distance without dissipation, whereas the latter accurately simulates the aerodynamics in the tip region.

Global performance calculations (radial distributions of lift) are validated by a comparison to the experimental database of Caradonna and Tung[1]. Good agreement is found for all pitch angles and rotational speeds, including transonic flow conditions.

Comparisons with the previously published methods of Komerath *et al.*[3], Srinivasan and McCroskey[4] and Vion *et al.*[26] indicate considerable improvement in the prediction of the blade aerodynamics.

The flow around the blade tip is investigated in detail by a comparison with the database of Gray *et al.*[16]. Good agreement is obtained with the hybrid CFD method, except for a slight difference in the vortex trajectory over the blade. This deviation is likely due to the absence of centrifugal effects and, to a lesser extent, of a radial component of induced velocity in the computation.

The low computational cost of the steady RANS method and the efficient and reliable hybrid adaptation method can now be used for further investigations, including variations of the blade geometry and tip shape.

## ACKNOWLEDGMENTS

The authors thank Michael Le Bars (IRPHE-CNRS) for discussions on the centrifugal and Coriolis forces.